\documentclass{article}

\usepackage{spconf}
\usepackage{amsmath,graphicx}
\usepackage{amssymb,amsmath,bm}
\usepackage{textcomp}
\usepackage{algorithm}
\usepackage{algpseudocode}
\usepackage{float}
\usepackage{multirow}
\usepackage{subcaption}
\usepackage[justification=centering]{caption}

\usepackage{tikz,placeins}
\tikzstyle{every picture}+=[font=\rmfamily\it\bfseries\large]

\graphicspath{{../figures/}}

\newcommand{\specialcell}[2][c]{%
     \begin{tabular}[#1]{@{}c@{}}#2\end{tabular}}

\DeclareMathOperator*{\argmax}{arg\,max}

\title{End-to-end Training of a Large Vocabulary \\ End-to-end Speech Recognition System}
\name{Chanwoo Kim, Sungsoo Kim, Kwangyoun Kim, Mehul Kumar,  Jiyeon Kim,
Kyungmin Lee, Changwoo Han, 
\thanks{Thanks to Samsung
Electronics for funding this research. The authors are thankful to  
Executive Vice President Seunghwan Cho, Ravichander Vipperla, Nicholas Lane,
and members of Speech Processing Lab
 at Samsung Research.}
}
\secondlinename{Abhinav Garg, Eunhyang Kim,  Minkyoo Shin, Shatrughan Singh,
Larry Heck, Dhananjaya Gowda}

\address{Samsung Research \\
{\small \tt \{chanw.com, ss216.kim, ky85.kim, mehul3.kumar, jstacey7.kim,
k.m.lee, cw1105.han} \\ \small \tt {abhinav.garg, sc.ehkim.jin, mk0211.shin, shatrughan.s,
larry.h, d.gowda\}@samsung.com }}

\begin{document}
\ninept
\maketitle
\begin{abstract}
In this paper, we present an end-to-end training framework for 
building state-of-the-art end-to-end speech recognition systems.
Our training system utilizes a cluster of Central Processing Units 
  (CPUs) and Graphics Processing Units (GPUs).
The entire data reading, large scale data augmentation,
neural network parameter updates are all performed ``on-the-fly". 
  We use vocal tract length perturbation \cite{c_kim_interspeech_2019_00} 
  and an acoustic simulator \cite{c_kim_interspeech_2017_00}
for data augmentation. The processed features and labels are sent 
to the GPU cluster. The Horovod {\tt allreduce} approach is employed to train
neural network parameters.
We evaluated the effectiveness of our system on the standard 
Librispeech corpus \cite{v_panayotov_icassp_2015_00} and 
  the 10,000-hr anonymized Bixby English dataset. 
Our end-to-end speech
recognition system built using this training infrastructure showed 
  a 2.44 \% WER on {\tt test-clean} of the LibriSpeech test set 
 after applying shallow fusion with a Transformer language model (LM).
 For the proprietary English Bixby open domain test set,
we obtained a WER of 7.92 \% using 
a Bidirectional Full Attention (BFA) end-to-end model after applying shallow fusion with an RNN-LM.
When the monotonic chunckwise attention (MoCha) based approach is employed for 
streaming speech recognition, we obtained a WER of 9.95 \% 
  on the same Bixby open domain test set.
\end{abstract}
%
\section{Introduction}
In recent years, deep learning techniques have significantly 
improved speech recognition accuracy \cite{Seltzer2013DNNAurora4, 
Yu2013FeatureLearningDNN, V_Vanhoucke_Deep_Learning_NIPS_Workshop_2011,
G_Hinton_IEEE_Signal_Process_Mag_2012,
T_Sainath_IEEETran_2017_1}.
This improvement has come about from the shift from Gaussian Mixture Model
(GMM) to the Feed-Forward Deep Neural Networks (FF-DNNs), FF-DNNs
to Recurrent Neural Network (RNN) and in particular the Long Short-Term Memory
(LSTM) networks \cite{S_Hochreiter_neural_computation_1997_00}. 
Thanks to these advances, voice assistant devices such as Google Home
\cite{c_kim_interspeech_2017_00, B_Li_INTERSPEECH_2017_1}
, Amazon Alexa or Samsung Bixby \cite{samsung_bixby} are being used at 
many homes and on personal devices. 

Recently there has been increasing interest in switching
from the conventional Weighted Finite State Transducer (WFST)
based decoder using an Acoustic Model (AM) and a Language Model (LM)
to a complete end-to-end all-neural speech recognition systems 
\cite{w_chan_icassp_2016_00, r_prabhavalkar_interspeech_2017_00, 
j_chorowski_nips_2015_00}. 
These complete end-to-end systems have started surpassing the performance of
the conventional WFST-based decoders with a very large training 
database \cite{c_chiu_icassp_2018_00}, a better choice of target unit 
such as Byte Pair Encoded (BPE) subword units, and an improved 
training methodology such as Minimum
Word Error Rate (MWER) training \cite{r_Prabhavalkar_icassp_2018_00}.

Another important aspect in building high-performance 
speech recognition systems is the amount
and the coverage of the training data.
To build high performance speech recognition systems for
conversational speech, we need to use
a large amount of speech data covering various domains
\cite{a_narayanan_slt_2018_00}. In \cite{h_soltau_interspeech_2017_00}, 
it has been shown that we need a very large training set ($\sim$125,000 hours of 
semi-supervised speech data) to achieve
high speech recognition accuracy for difficult tasks like 
video captioning. To train neural networks using such large 
amounts of speech data, we usually need 
multiple Central Processing Units (CPUs) or Graphics
Processing Units (GPUs)
\cite{E_Variani_INTERSPEECH_2017_01, p_goyal_arxiv_2018_00}.

With widespread adoption of voice assistant speakers, far-field
speech recognition has become very important.
In far-field speech recognition, the impacts of reverberation and noise 
are much larger than those in near-field cases. 
Traditional approaches to far-field
speech recognition include noise robust feature extraction algorithms
\cite{C_Kim_IEEETran_2016_1, U_H_Yapanel_SpeechComm_2008, c_kim_interspeech_2014_00},
or multi-microphone approaches
\cite{T_Nekatani_ICASSP_2017_1, T_Higuchi_ICASSP_2016_1,
H_Erdogan_INTERSPEECH_2016_1,
C_Kim_INTERSPEECH_2010_1, C_Kim_ICASSP_2012_2, C_Kim_INTERSPEECH_2015_1, c_kim_icassp_2018_01}.
More recently, approaches using data augmentation
has been gaining popularity for far-field speech recognition
\cite{R_Lippmann_icassp_1987_1,
c_kim_icassp_2018_00, w_hartmann_interspeech_2016_00, 
s_park_interspeech_2019_00, C_Kim_ASRU_2009_2}. 
An ``acoustic simulator"  
\cite{c_kim_interspeech_2017_00, c_kim_interspeech_2018_00}
is used to generate simulated speech utterances for
millions of different room dimensions, a wide distribution
of reverberation time and signal-to-noise ratio. In a similar 
spirit, Vocal Tract Length Perturbation (VTLP) has been proposed 
\cite{n_jaitly_icml_workshop_2013_00} to tackle the speaker variability issue.
As shown in our recent paper \cite{c_kim_interspeech_2019_00}, 
VTLP is especially useful when
the speaker variability in the training database is not sufficient. 
For these kinds of data augmentation, processing on CPUs is more
desirable than processing on GPUs. Due to this, we have proposed
an end-to-end training approach using Example Servers (ES)
 and workers. Example servers are typically run on the CPU
cluster performing data reading, data augmentation, and feature extraction
\cite{E_Variani_INTERSPEECH_2017_01, c_kim_interspeech_2018_00}.
\begin{figure*}[!tbp]
  \begin{center}
    \resizebox{\textwidth}{!}{\input{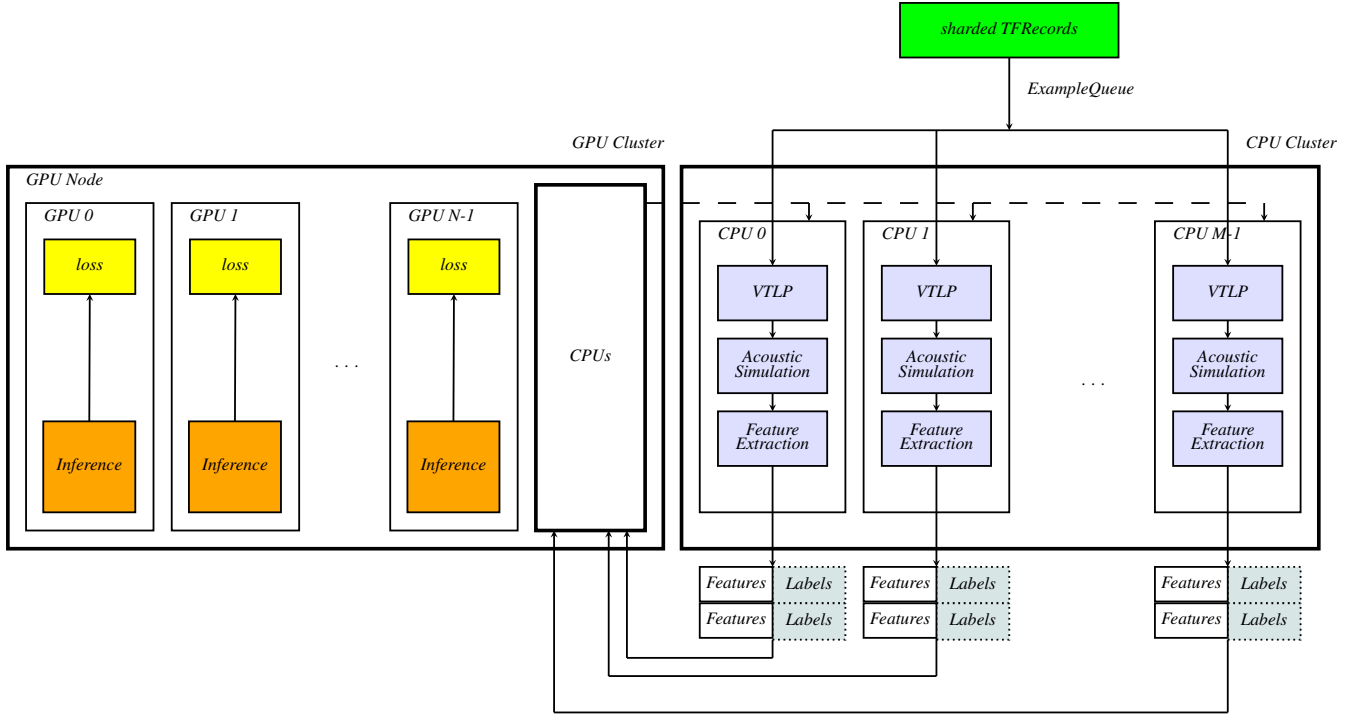}}
      \caption { The Samsung Research end-to-end training framework for building an end-to-end
	speech recognition system with multi CPU-GPU clusters and on-the-fly data
	processing and augmentation pipeline.
     }   
     \label{fig:end_to_end_training_system}
  \end{center}
\vspace{-5mm}
\end{figure*}
In this paper, we describe the structure of our end-to-end training system
to train an end-to-end speech recognition system. This training system
has several advantages over previous systems described in
\cite{c_kim_interspeech_2018_00}. First, instead of using the {\tt QueueRunner},
we use a more efficient data queue using {\tt tf.data} in Tensorflow 
\cite{m_abadi_usenix_2016}.  Second, instead
of pre-calculating information about room configurations and 
room impulse responses in the acoustic simulator, these are
calculated {\it on-the-fly}. Thus, the entire training system runs 
{\it on-the-fly}. Additionally, instead of using the parameter server-worker
structure, we use an {\tt allreduce} approach implemented in the Horovod
\cite{a_sergeev_arxiv_2018_00} distributed training framework, which
has been shown to be more efficient. The system described in 
\cite{E_Variani_INTERSPEECH_2017_01}, is designed
to train the acoustic model part of the speech recognition system where
as our training system trains the complete end-to-end speech recognition system.

The rest of the paper is organized as follows: We describe the entire
training system structure in detail in Sec. \ref{sec:training_cpu_gpu}.
The structure of the end-to-end speech recognition system 
is described in Sec. \ref{sec:end_to_end_speech_recognition}. 
Experimental results that demonstrates the effectiveness of our speech
recognition system is presented in Sec. \ref{sec:experimental_results}. 
We conclude in Sec.  \ref{sec:conclusions}.
\section{Overall structure of the end-to-end speech recognition}
\label{sec:training_cpu_gpu}
\vspace{-1mm}

In this section, we describe the overall structure of our 
end-to-end training system.  Fig. \ref{fig:end_to_end_training_system}
shows how the entire system is structured. Our system consists
of a cluster of CPUs and a cluster of GPUs. Each GPU node of 
the GPU cluster has eight Nvidia\texttrademark P-40, P-100 or V-100 GPUs and two 
Intel E5-2690 v4 CPUs. Each of these CPUs has 14 cores.
The large box on the left hand side of Fig. \ref{fig:end_to_end_training_system}
denoted ``GPU cluster'' shows a typical GPU node with $N$ GPUs.
The large box on the right shows a ``CPU cluster'' of $M$ CPUs,
each running an independent data pipeline. 

\vspace{-1mm}
\subsection{Training job launch}
The main process of the training system runs on one of CPU cores 
of the GPU cluster. This CPU portion of the GPU node is represented
as a box in the right hand side of the GPU node box.
When the training job starts, this main training process launches
multiple example server jobs on the CPU cluster using the 
{\tt IBM Platform LSF} \cite{ibm_spectrum_lsf_2010}. In Fig. 
\ref{fig:end_to_end_training_system}, this launching process is represented
by a dashed arrow from the CPU portion of the GPU node to the 
CPU cluster.

\subsection{Data reading using an example queue}
In the CPU cluster, each CPU runs one example server
which reads speech utterance and transcript data from sharded 
{\tt TFRecords} defined in Tensorflow \cite{m_abadi_usenix_2016}.
The TFRecord format is a simple format in Tensorflow for storing 
a sequence of binary records. To support efficient reading using 
multiple CPUs, we use sharded {\tt TFRecords}.

To read large-scale data efficiently in parallel, we use an example queue
shown in the left side of Fig. \ref{fig:end_to_end_training_system}.  
The original speech waveform data, transcripts, and meta data are stored
in sharded {\tt TFRecords}. The data pipeline is implemented using
{\tt tf.data} in Tensorflow \cite{m_abadi_usenix_2016}, 
and contains the data augmentation and feature extraction blocks.
These {\tt tf.data} APIs are efficient in building complex pipelines
by applying a series of elementary operations. We perform data interleaving
and parallel reading using {\tt tf.contrib.data.parallel\_interleave}, 
shuffling using {\tt tf.data.Datatset.shuffle}, and padding using 
{\tt tf.data.Dataset.padded\_batch}.

\subsection{Data augmentation and feature extraction}
\label{sec:feature_extraction}
To improve robustness against speaker variability, we apply an
on-the-fly VTLP algorithm on the input waveform
\cite{c_kim_interspeech_2019_00}. The warping
factor is generated randomly for each input utterance. Unlike
conventional VTLP approaches in \cite{n_jaitly_icml_workshop_2013_00,   
x_cui_taslp_2015_00}, we resynthesize the processed
speech. The purpose of doing this is to apply VTLP before applying
the acoustic simulator to the input waveform.  This is quite 
natural that data augmentation to model speaker variability 
should be performed before the data augmentation to model
acoustic variability.
One more advantage is that this
resynthesis approach enables us to use a window length 
optimal for VTLP different from that used in feature processing.
We apply a blinear transformation \cite{p_zhan_cmu_tech_report_1997_00} 
to perform frequency warping
to model speaker variability due to the difference in the vocal tract length.
In the bilinear transformation, the relation between the input
and output discrete-time frequencies is given by: 
\begin{align}
  \omega_k'  = \omega_k + 2 \tan^{-1} \left(
                    \frac{ \left(1 - \alpha \right) \sin(\omega_k)} 
              {1 - (1 - \alpha) \cos(\omega_k) } \right)
                            \label{eq:bilinear_transformation}.
\end{align}
where $\omega_k = \frac{2 \pi k}{K}$ is the input discrete-time frequency
, $\omega_k' = \frac{2 \pi k'}{K}$ is the output discrete-time frequency, and
  $K$ is the DFT size. 
More details about our VTLP algorithm is described 
in detail in \cite{c_kim_interspeech_2019_00}.
The acoustic simulator in Fig.~\ref{fig:end_to_end_training_system}
is similar to what we described in 
\cite{c_kim_interspeech_2017_00, c_kim_interspeech_2018_00}.
One difference compared to our previous one in 
\cite{c_kim_interspeech_2017_00} is that we do not pre-calculate
room impulse responses, but instead they are calculated on-the-fly.
For feature processing we use {\tt tf.data.Dataset.map} API. 
Instead of using the more conventional log-mel or 
MFCC features, we use the power mel filterbank energies,
since it shows slightly better performance \cite{c_kim_interspeech_2019_00,
c_kim_asru_2019_00}. Motivated by our previous research of
using power-law nonlinearity with a power coefficient between 
$\frac{1}{15}$ \cite{C_Kim_ICASSP_2010_1, C_Kim_ICASSP_2012_1, C_Kim_PhDThesis_2010} and 
$\frac{1}{10}$ \cite{C_Kim_INTERSPEECH_2009_2}, we apply the 
power-law nonlinearity of $(\cdot)^{\frac{1}{15}}$ to the mel filterbank 
coefficients. We refer to this feature as {\it power-mel filterbank
coefficients}.

\begin{figure}[tbp]
  \begin{center}
    \resizebox{80mm}{!}{\input{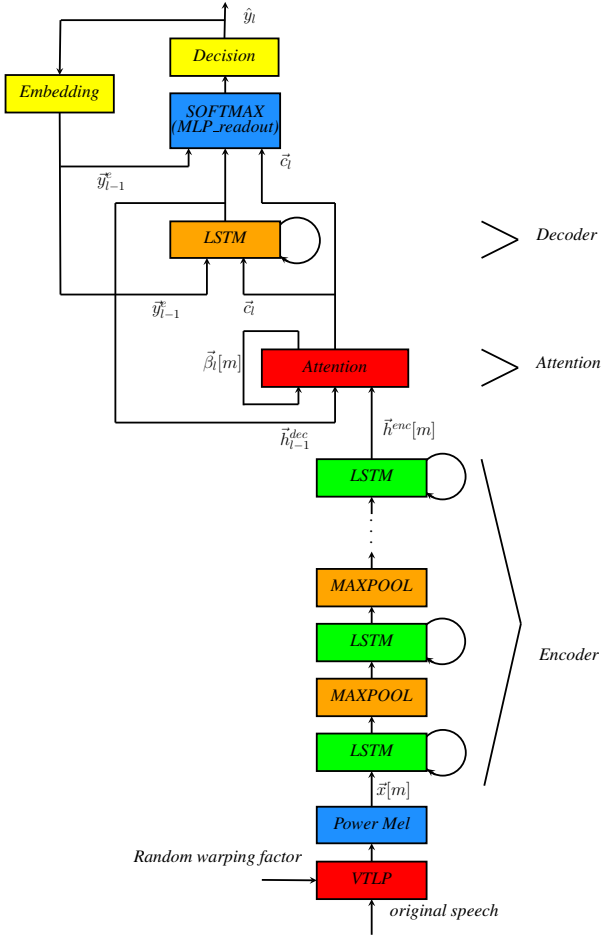}}
      \caption {  The structure of the entire end-to-end
      speech recognition system. 
     }
     \label{fig:entire_diagram}
  \end{center}
\vspace{-5mm}
\end{figure}

\subsection{Parameter calculation and update}
\label{sec:parameter_update}
The features and the target labels are sent to the GPU cluster
using the ZeroMQ \cite{zero_mq} asynchronous messaging queue.
Each {\it example server } sends these data asynchronously to the CPU portion
of the GPU node as shown in Fig \ref{fig:end_to_end_training_system}.
Using these data, neural network parameters are calculated and updated 
using an Adam optimizer and the
Horovod \cite{a_sergeev_arxiv_2018_00} {\tt allreduce} approach.

\begin{figure}
  \captionsetup[subfigure]{justification=centering}
    \centering
    \begin{subfigure}[5]{0.5\textwidth}
      {\includegraphics[width=80mm]{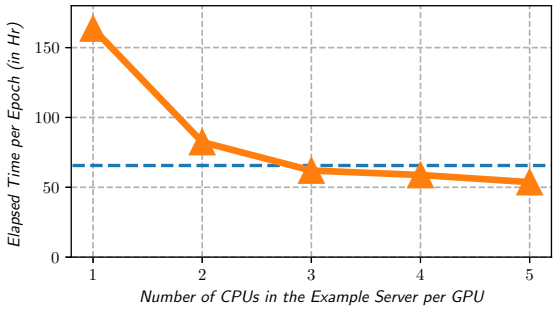}}
      \caption{
          \label{fig:plot_example_server_time}
      }
    \end{subfigure}

    \begin{subfigure}[b]{0.5\textwidth} 
      {\includegraphics[width=80mm]{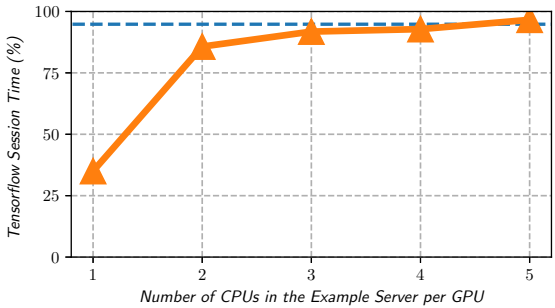}}
    \caption{
      \label{fig:plot_gpu_time_percentage}
    }
    \end{subfigure}
  \caption{\label{fig:plot_example_server_performance}
  The efficiency of the {\it example server} with respect to
  the number of CPUs per GPU: (a) The required time to 
  process a single epoch during the training phase and (b) 
  the percentage of {\tt Tensorflow } computation time defined by 
  \eqref{eq:session_time_portion}. 
  The blue horizontal dotted lines in each figure represent the case
  when data augmentation with example servers is not employed.
  }
\end{figure}

Fig. \ref{fig:plot_example_server_performance} shows how many CPUs
in the {\it  example server } per GPU are required to provide sufficient 
data to the GPU cluster. In this experiment, we used a 10,000-hr
anonymized {\texttt Bixby} English training set. 
We trained a streaming end-to-end model using the Monotonic CHunkwise
Attention (MoCha) algorithm \cite{c_chiu_iclr_2018_00}. The details
about our MoCha implemention are discussed in \cite{k_kim_asru_2019_00}.
In the {\it  example server}, we ran the
VTLP data augmentation \cite{c_kim_interspeech_2019_00}, 
{\it acoustic simulator} \cite{c_kim_interspeech_2017_00} and 
feature extraction modules shown in Fig. \ref{fig:entire_diagram}.
In this experiment, we used four Nvidia\texttrademark V-100 GPUs with 32-GB
memory in the 
GPU cluster. Fig. \ref{fig:plot_example_server_time} shows how much
time is required to finish one epoch of training. When data augmentation
is not applied, 65.6-hours were required to finish one epoch of training.
Fig. \ref{fig:plot_example_server_time} shows us that three CPUs per GPU 
(total 12 CPUs for 4 GPUs) are required to obtain a comparable throughput. 
If we use four or five CPUs per GPU, as shown in Fig.
\ref{fig:plot_example_server_time}, the training is even slightly 
faster than the case without the {\it  example-server}-based data-augmentation.
We think that this happened because of more efficient data processing
with the {\it example server}. When we do not perform data augmentation
using {\it example servers}, feature extraction and data reading are performed
on a limited numbers of CPUs inside the GPU cluster, which might add some
latency during the training. 
Thus, it is possible that the training with data augmentation
using {\it example servers} may be even slightly faster than the 
baseline case without data-augmentation using {\it example servers}.

Fig. \ref{fig:plot_gpu_time_percentage} shows the portion of the time
used for{\tt Tensorflow } computation. This portion of time is defined
by:
\begin{align}
  t_{\text{session}} =  \frac{\text{Time Spent in {\tt Tensorflow}
  Session}}{\text{Elapsed Time}}.
  \label{eq:session_time_portion}
\end{align}
If GPUs in the GPU cluster are not given sufficient amount 
of data, these GPUs will remain idle. Thus, $t_{\text{session}}$ in
\eqref{eq:session_time_portion} is a good indicator to 
see whether the {\it example server } 
provides sufficient 
amount of processed features. From this figure, we may conclude that
three $\sim$ four CPUs per GPU (total 12 $\sim$ 16 CPUs for 4 GPUs) 
are required to keep GPUs busy enough. In our experiments using
the 10,000-hr Bixby training set in Sec. \ref{sec:experimental_results},
we used 8 GPUs and 40 CPUs (5 CPUs per GPU) during the training.

\section{Structure of the end-to-end speech recognition system}
\label{sec:end_to_end_speech_recognition}

We have adopted the RETURNN speech recognition system \cite{p_doetsch_icassp_2017_00,
a_zeyer_interspeech_2018_00} for training our end-to-end system with various modifications.
Some of the important modifications are: replacing the input data pipeline with our 
proposed on-the-fly example server based pipeline with support for VTLP
and acoustic simulation,
implementing the Monotonic Chunkwise Attention (MoChA) \cite{c_chiu_iclr_2018_00}
for online streaming end-to-end speech recognition, minimum Word Error Rate (mWER) 
training, support for handling Korean language or script, our own scoring and Inverse Text 
Normalization (ITN) modules, support for power mel filterbank features
\cite{c_kim_interspeech_2019_00, c_kim_asru_2019_00}, etc. We have tried
various types of training strategies for better performance
\cite{d_gowda_interspeech_2019_00, a_garg_asru_2019_00}.
Our MoCha implementation and optimization are described in very detail in our
another paper \cite{k_kim_asru_2019_00}.

The structure of our entire end-to-end speech recognition system
 is shown in Fig.~\ref{fig:entire_diagram}.
 $\vec{x}[m]$ and $\vec{y}_l$ are the input power mel 
filterbank energy vector and the output label,
respectively. $m$ is the input frame index and $l$ is the decoder output
step index. $\vec{c}_l$ is the context vector calculated 
as a weighted sum of the encoder hidden state vectors denoted as $\vec{h}^{enc}[m]$.
The attention weights are computed as a softmax of energies computed as 
a function of the encoder hidden state $\vec{h}^{enc}[m]$, the decoder hidden state
$\vec{h}^{dec}_l$, and the attention weight
feedback $\vec{\beta}_l[m]$  \cite{a_zeyer_interspeech_2018_00}.

In \cite{a_zeyer_interspeech_2018_00}, the peak value of the speech 
waveform is normalized to be one. However, since finding the peak sample value 
is not possible for online feature extraction, we do not perform this
normalization. We modified the input pipeline
so that the online feature generation can be performed. We disabled the 
clipping of feature range between -3 and 3, which is the default 
setting for the Librispeech experiment using MFCC 
features in \cite{a_zeyer_interspeech_2018_00}. 
We conducted experiments using both the 
uni-directional and bi-directional Long Short-Term Memories (LSTMs)
\cite{S_Hochreiter_neural_computation_1997_00} in the encoder.
However, only the uni-directional LSTMs are used in the decoder.
For online speech recognition experiments, we used the
MoChA models \cite{c_chiu_iclr_2018_00} with a chunk size of 2.
In MoCha experiments, we used only the uni-directional LSTMs 
both in the encoder and the decoder to enable streaming recognition.
 For better stability in LSTM training, we use the gradient clipping by  
global norm \cite{r_pascanu_icml_2013}, which is implemented as  
{\tt tf.clip\_by\_global\_norm } API in Tensorflow  \cite{m_abadi_usenix_2016}.
We use six layers of encoders and one layer of decoder followed by a softmax
layer.

In performing shallow-fusion with an external LM, our approach is slightly 
different from the previously known approaches 
\cite{c_gulcehre_corr_2015_00, s_toshniwal_slt_2018_00} to obtain better 
performance.
we used the following equation:
\begin{align}
  y_{0:L}^{*} = & \argmax_{y_{0:L}} \sum_{l=0}^{L-1} \Big[ \log P \left(y_l
  |\vec{x}[0:M], y_{0:l} \right)
  \nonumber \\
                           & \qquad - \lambda_p \log P(y_l) 
                           + \lambda_{\text{lm}} \log P \left(y_l   |y_{0:l} \right)  \Big],
                           \label{eq:lm_fusion}
\end{align}
where we have an additional term $\lambda_p \log P(y_l) $ for subtracting 
the {\it prior bias} that the model has learned from the training corpus. 
In \eqref{eq:lm_fusion} $L$ is the 
length of the output label hypothesis. $\lambda_p$ and $\lambda_{\text{lm}}$ are 
weights for the prior probability and the LM prediction probability, 
respectively.
In \eqref{eq:lm_fusion}, we represented sequences following the Python slice
notation. For example, $\vec{x}[0:M]$ denotes the sequence of the input acoustic
features of length $M$, and $y_{0:L}$ is a sequence of output labels  of length $L$.
\section{Experimental Results}
\label{sec:experimental_results}
In this section, we present a summary of experimental results obtained
with our end-to-end speech recognition systems built using the proposed
Samsung Research end-to-end training framework.
For near-field speech recognition experiments, we use the open source 
Librispeech database \cite{v_panayotov_icassp_2015_00}, as well as our in-house
{\it Bixby} \cite{samsung_bixby} usage training and test sets for English. 
The LibriSpeech dataset consists of around 960 hours of training data
consisting of 281,241 utterances. The evaluation set consists of the 
official 5.4 hours {\tt test-clean} and 5.1 hours {\tt test-other} data.
The Bixby training set consists of approximately 10,000 hours of
anonymized Bixby usage data. The evaluation set consists of around 1,000 
open domain utterances. As mentioned in Sec. 
\ref{sec:parameter_update}, we used 8 GPUs in a GPU cluster and 
40 CPUs in an example server when training the model using the 
Bixby training set.
\begin{table}[!tbhp]
  \renewcommand{\arraystretch}{1.3}
  \centering
        \caption{\label{tbl:power_law_result}
        Word Error Rates (WERs) obtained using MFCC implemented in
        \cite{b_mcfee_proc_scipy_2015_00} and power mel filterbank coefficients
        \\ on the Librispeech corpus \cite{v_panayotov_icassp_2015_00}. For
        each WER number, \\the same experiment was conducted twice and averaged.
        }
        \begin{tabular}{| c | c  || c | c | c | c |}
          \hline
          \multicolumn{2}{| l ||}{Cell Size}
                                 & \specialcell{MFCC}
                                 & \specialcell{Power Mel \\ 
                                 Filterbank \\
                                 Coefficients} \\
          \hline
          \multirow{3}{*}{1536 cell}    
                  & test-clean  &   4.06  \% &  \textcolor{blue}{\bf 3.94 } \%   \\
                  & test-other  &  13.97  \% &  \textcolor{blue}{\bf 13.56} \%   \\
                  & average     &   9.02  \% &  \textcolor{blue}{\bf 8.75 } \%   \\
          \hline
       \end{tabular}
       \vspace{-2mm}
\end{table}
\begin{table}[!tbhp]
  \renewcommand{\arraystretch}{1.3}
  \centering
        \caption{\label{tbl:vtlp_result}
        Word Error Rates (WERs) obtained with VTLP processing
        with different warping factor $\alpha$ distribution, 
        and with and without an RNN LM. The warping factor $\alpha$
			  is the constant controlling warping in \eqref{eq:bilinear_transformation}.
        }
        \begin{tabular}{| c | c  || c | c | c | c |}
          \hline
          \multicolumn{2}{| l ||}{Warping Factor}
                                 & \specialcell{ 0.7 $\sim$ 1.3 }
                                 & \specialcell{ 0.8 $\sim$ 1.2 }  
                                 & \specialcell{ 0.9 $\sim$ 1.1 }   \\
          \hline \hline
          \multirow{3}{*}{\specialcell{Without \\ RNN-LM}}
                  & test-clean  &   3.82  \% &  3.66  \%   &  3.86  \%  \\  
                  & test-other  &  12.50  \% & 12.39  \%   & 12.35  \%  \\  
                  & average     &   8.16  \% &  8.03  \%   &  8.11  \%  \\  
          \hline
          \multirow{3}{*}{\specialcell{With \\ RNN-LM}}
                  & test-clean  &   2.93  \% &  \textcolor{blue}{\bf 2.85}  \%   &  2.96  \%  \\  
                  & test-other  &  10.40  \% & 10.25  \% & \textcolor{blue}{\bf10.13}  \%  \\  
                  & average     &   6.67  \% &  \textcolor{blue} {\bf 6.55}  \%
                  &  \textcolor{blue}{\bf 6.55}  \%  \\  
          \hline
       \end{tabular}
       \vspace{-2mm}
\end{table}

In Table \ref{tbl:power_law_result}, we compare the performance between
the baseline MFCC and the power-law of $(\cdot)^{\frac{1}{15}}$ features 
for a Bidirectional Full Attention (BFA) end-to-end model with 
an LSTM cell size of 1536 on the LibriSpeech database \cite{v_panayotov_icassp_2015_00}. 
Especially for {\tt test-other}, which
is a more difficult task, the power mel filterbank coefficients 
shows better performance than the baseline MFCC. Thus, we use 
the power-mel filterbank coefficients as the default feature in 
our end-to-end system. All the following results in this section
were obtained using the power-mel filterbank coefficients.

In Table \ref{tbl:vtlp_result}, we show Word Error Rates (WERs) on the same
LibriSpeech corpus for a BFA model using 
different window sizes and warping coefficient distributions, with and without
using an external Recurrent Neural Network (RNN) Language Model (LM) 
\cite{a_zeyer_interspeech_2018_00} built using the standard LibriSpeech LM corpus.
The best performance was achieved when the window length is 50 {\it ms} and 
the warping coefficients are uniformly distributed between 0.8 and 1.2.
We obtained 3.66 \% WER on the {\tt test-clean} database and 12.39 \%
WER on the {\tt test-other} database without using an LM. 
Using this shallow-fusion technique 
  with an RNN-LM, we achieved WERs of 2.85 \% and 10.25 \% 
  on the Librispeech {\tt test-clean} and {\tt test-other} databases, respectively.

\begin{table}[!htbp]
  \renewcommand{\arraystretch}{1.3}
  \centering
        \caption{\label{tbl:vtlp_result_with_trans_lm}
        Word Error Rates (WERs) obtained with VTLP processing
        using shallow-fusion with a Transformer LM with different beam sizes.
        The window length is 50 ms, and the warping factor distribution is
        $0.8 \sim 1.2$.
        }
        \begin{tabular}{| c || c | c | c | c |}
          \hline
                                 {Beam Size}
                                 & \specialcell{ 12 }
                                 & \specialcell{ 24 }   
                                 & \specialcell{ 36 }   
                                 & \specialcell{ 48 }   \\  
          \hline \hline
              \specialcell{ $\lambda_p$ \\ $ \lambda_{\text{lm}}$ } &   
              \specialcell{ 0.005\\ 0.45}  &
              \specialcell{ 0.004\\ 0.46}  &   
              \specialcell{ 0.003\\ 0.48}  &   
              \specialcell{ 0.002\\ 0.48} \\
              \hline
    
              test-clean  &   2.49  \% & 2.45 \%  &  \textcolor{blue}{\bf 2.44 \%} & 2.45 \%  \\  
              test-other  &   8.76  \% & 8.40 \%  &   8.29 \%
              & \textcolor{blue}{\bf 8.22
              \%} \\
              average     &   5.63  \% & 5.43  \% &   5.37 \%
              & \textcolor{blue}{\bf 5.34 \%} \\
          \hline
       \end{tabular}
       \vspace{-2mm}
\end{table}
\begin{figure}
  \captionsetup[subfigure]{justification=centering}
    \centering
    \begin{subfigure}[5]{0.5\textwidth}
      {\includegraphics[width=75mm]{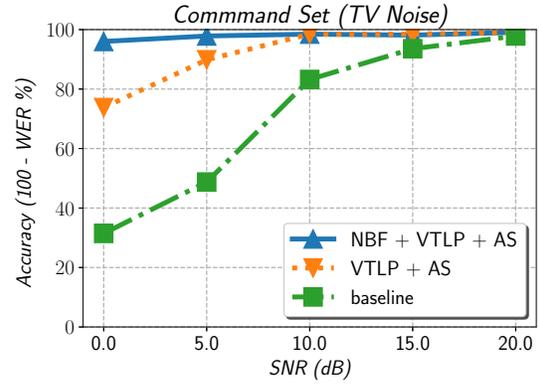}}
      \caption{
          \label{fig:tv_noise}
      }
    \end{subfigure}

    \begin{subfigure}[b]{0.5\textwidth} 
      {\includegraphics[width=75mm]{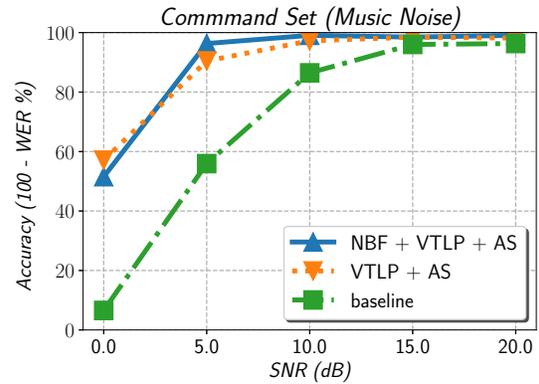}}

    \caption{
          \label{fig:music_noise}
    }
    \end{subfigure}

    \begin{subfigure}[b]{0.5\textwidth} 
      {\includegraphics[width=75mm]{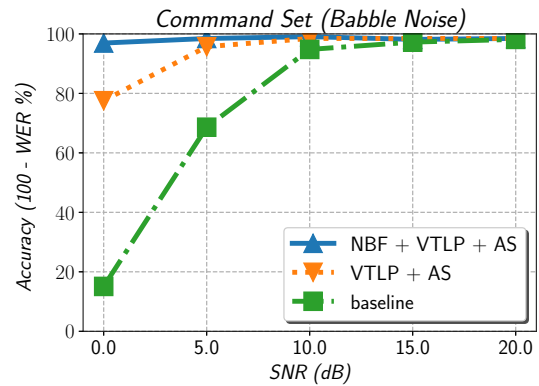}}
    \caption{
          \label{fig:babble_noise}
    }
    \end{subfigure}

 \caption{\label{fig:plot_farfield_wer}
  Speech recognition accuracy at different Signal-to-Noise Ratios (SNRs)
  under three different noisy conditions: direct TV noise (a)
  , music noise (b), and babble noise (c). NBF, VTLP, and AS
  stand for Neural Beam Former (NBF) \cite{j_heymann_icassp_2016_00}, 
  Vocal Tract Length Perturbation (VTLP) \cite{c_kim_interspeech_2019_00} , and
  Acoustics Simulator (AS) \cite{c_kim_interspeech_2017_00,
  c_kim_interspeech_2018_00},
  respectively.
  \vspace{-2mm}
 }
\end{figure}
Table \ref{tbl:vtlp_result_with_trans_lm} shows word error rates on the
  LibriSpeech testsets obtained 
  by applying shallow-fusion with a Transformer LM 
  \cite{a_vaswani_nips_2017_00, c_luscher_arxiv_2019_00} using
  \eqref{eq:lm_fusion}. As shown in this table, we conducted experiments with
  different beam sizes, $\lambda_p$ and $\lambda_{\text{lm}}$ parameters in
  \eqref{eq:lm_fusion}. The best result we obtained using a Transformer LM
  in Table \ref{tbl:vtlp_result_with_trans_lm} is significantly better than the result we obtained with a LSTM LM in 
  Table \ref{tbl:vtlp_result}.

In Table \ref{tbl:near_field_result}, we summarize our WER results for 
both the LibriSpeech and Bixby end-to-end ASR models.
In the case of the Bixby model, we optionally used an external RNN-LM 
trained using around 65GB of 
the Bixby LM corpus with an architecture exactly similar to the
LibriSpeech LM model used in \cite{a_zeyer_interspeech_2018_00}.
The cell sizes of the LibriSpeech model and the Bixby model 
in Table \ref{tbl:near_field_result} are 1536 and 1024 respectively.
For comparison, the best WFST based conventional LSTM-HMM based ASR system
gives a WER of 8.85\% on the Bixby same open domain test set.
We can see that our current Bixby end-to-end BFA model is $\sim$10\% better, while
our MoChA streaming model is $\sim$10\% poorer compared to
the conventional WFST based DNN-HMM system.
\begin{table}[!tbhp]
  \renewcommand{\arraystretch}{1.3}
  \centering
        \caption{\label{tbl:near_field_result}
        Summary of Word Error Rates (WERs) obtained for different
        LibriSpeech and Bixby near-field end-to-end ASR models 
        with and without an RNN LM.
        }
        \begin{tabular}{| c | c  || c | c |}
          \hline
          \multicolumn{2}{| l ||}{Models}
                                 & BFA
                                 & MoChA \\
          \hline \hline
          \multirow{3}{*}{\specialcell{LibriSpeech \\ (1536-cell) \\ test-clean}}
                  & w/o LM  &   3.66  \% &  6.78  \%  \\  
                  & RNN-LM  &   2.85  \% &  5.54  \%  \\  
                  & Transformer LM  &   2.44  \% &   -   \\  
          \hline
          \multirow{2}{*}{\specialcell{Bixby \\ (1024-cell) }}
                  & w/o LM  &   8.25  \% &  10.77  \% \\  
                  & RNN-LM  &   7.92  \% &   9.95  \% \\  
          \hline
       \end{tabular}
       \vspace{-2mm}
\end{table}
The performance of our far-field end-to-end ASR model trained using the proposed
structure in Fig. \ref{fig:entire_diagram} is shown 
in Fig. \ref{fig:plot_farfield_wer}. In this experiment, we used
the same anonymized 10,000 hours of the English Bixby training set.
The performance of the far-field
models were evaluated on the English Bixby command test set.
The Bixby command test set
include utterancesare sampled from the anonymized Bixby usage log. 
Examples in this
test set include ``Set an alarm for tomorrow at 6 a.m", 
``Tell me remaining time of my timers", ``Play the most latest added 
song", and so on.

Far-field recording using this Bixby command set was performed
by playing back utterances using a loud speaker at 5-meter distance 
in a real room.
The reverberation time in this
recording room was measured to be $T_{60}=430 ms$. We used two microphones
on a prototype {\it Galaxy Home Mini} to record this far-field speech.
The distance between two microphones is 6.8 cm. We simulated 
far-field additive noise by playing back three different types of noise using loud speakers.
In Fig. \ref{fig:tv_noise}, we used a single loud speaker located at 
1-meter distance from the microphone to simulate direct noise from 
a television. In Figs. \ref{fig:music_noise} and \ref{fig:babble_noise},
we used four loud speakers oriented to different directions to simulate
diffuse noise. On the prototype {\it Galaxy home Mini} device, 
the two-microphone signals are enhanced using a beamformer 
based on the Neural Network supported
Generalized Eigenvalue (NN-GEV) algorithm \cite{j_heymann_icassp_2016_00}. 
In Fig.  \ref{fig:plot_farfield_wer}, we evaluated speech recognition accuracy
using three different systems. {\tt NBP+VTLP+AS} denotes a system
which uses the VTLP system and the acoustic simulator described in this paper for 
data augmentation in model training, and additionally uses this NN-GEV-based 
beamformer for signal enhancement. {\it NBF} in this future stands for
this Neural Beam Former (NBF).  {\tt VTLP+AS} denotes a system employing
the VTLP system and the acoustic simulator without using this beamformer.
{\tt baseline} denotes a system which was trained using utterances
recorded in close-talking environments without any further processing. 
As can be seen in these figures,
data augmentation technique significantly enhances speech recognition
accuracy under the far-field environments. We also observe that
the data augmentation algorithm described in Sec. \ref{sec:training_cpu_gpu}
does not harm the clean performance, thus we may use the same 
data augmentation both for the close-talking and the far-field environments.
In case of the direct
TV noise in Fig. \ref{fig:tv_noise} and babble noise in Fig.
\ref{fig:babble_noise}, we observe that further improvement is 
achieved by employing a NN-GEV-based beamformer.
\section{Conclusions}
\label{sec:conclusions}
We presented a new end-to-end training framework and
 strategies for training 
state-of-the-art end-to-end speech recognition systems.
Our training system utilizes a cluster of Central Processing Units 
  (CPUs) and Graphics Processing Units (GPUs).
The entire data reading, large scale data augmentation,
neural network parameter updates are performed on-the-fly
using example servers and sharded {\tt TFRecords} and {\tt tf.data}. 
  We use vocal tract length perturbation and an acoustic simulator 
for data augmentation. Horovod {\tt allreduce} approach is employed 
to train the neural network parameters using Adam optimizer.
We evaluated the effectiveness of our system on the standard 
Librispeech corpus \cite{v_panayotov_icassp_2015_00} and 
  10,000-hr anonymized Bixby English training 
 and test sets both in near-field as well as
far-field scenarios.  
\bibliographystyle{IEEEtran}
\bibliography{common_bib_file}

\begin{thebibliography}{10}
\providecommand{\url}[1]{#1}
\csname url@samestyle\endcsname
\providecommand{\newblock}{\relax}
\providecommand{\bibinfo}[2]{#2}
\providecommand{\BIBentrySTDinterwordspacing}{\spaceskip=0pt\relax}
\providecommand{\BIBentryALTinterwordstretchfactor}{4}
\providecommand{\BIBentryALTinterwordspacing}{\spaceskip=\fontdimen2\font plus
\BIBentryALTinterwordstretchfactor\fontdimen3\font minus
  \fontdimen4\font\relax}
\providecommand{\BIBforeignlanguage}[2]{{%
\expandafter\ifx\csname l@#1\endcsname\relax
\typeout{** WARNING: IEEEtran.bst: No hyphenation pattern has been}%
\typeout{** loaded for the language `#1'. Using the pattern for}%
\typeout{** the default language instead.}%
\else
\language=\csname l@#1\endcsname
\fi
#2}}
\providecommand{\BIBdecl}{\relax}
\BIBdecl

\bibitem{c_kim_interspeech_2019_00}
\BIBentryALTinterwordspacing
{\chanwcom}, M.~Shin, A.~Garg, and D.~Gowda, ``{Improved Vocal Tract Length
  Perturbation for a State-of-the-Art End-to-End Speech Recognition System},''
  in \emph{INTERSPEECH-2019}, Graz, Austria, Sept. 2019, pp. 739--743.
  [Online]. Available: \url{http://dx.doi.org/10.21437/Interspeech.2019-3227}
\BIBentrySTDinterwordspacing

\bibitem{c_kim_interspeech_2017_00}
\BIBentryALTinterwordspacing
{\chanwcom}, A.~Misra, K.~Chin, T.~Hughes, A.~Narayanan, T.~N. Sainath, and
  M.~Bacchiani, ``Generation of large-scale simulated utterances in virtual
  rooms to train deep-neural networks for far-field speech recognition in
  google home,'' in \emph{Proc. Interspeech 2017}, 2017, pp. 379--383.
  [Online]. Available: \url{http://dx.doi.org/10.21437/Interspeech.2017-1510}
\BIBentrySTDinterwordspacing

\bibitem{v_panayotov_icassp_2015_00}
V.~Panayotov, G.~Chen, D.~Povey, and S.~Khudanpur, ``Librispeech: An asr corpus
  based on public domain audio books,'' in \emph{IEEE Int. Conf. Acoust.,
  Speech, and Signal Processing}, April 2015, pp. 5206--5210.

\bibitem{Seltzer2013DNNAurora4}
{M. Seltzer, D. Yu, and Y.-Q. Wang}, ``An investigation of deep neural networks
  for noise robust speech recognition,'' in \emph{Int. Conf. Acoust. Speech,
  and Signal Processing}, 2013, pp. 7398--7402.

\bibitem{Yu2013FeatureLearningDNN}
{D. Yu, M. L. Seltzer, J. Li, J.-T. Huang, and F. Seide}, ``Feature learning in
  deep neural networks - studies on speech recognition tasks,'' in
  \emph{Proceedings of the International Conference on Learning
  Representations}, 2013.

\bibitem{V_Vanhoucke_Deep_Learning_NIPS_Workshop_2011}
{V. Vanhoucke, A. Senior, and M. Z. Mao}, ``{Improving the speed of neural
  networks on CPUs},'' in \emph{Deep Learning and Unsupervised Feature Learning
  NIPS Workshop}, 2011.

\bibitem{G_Hinton_IEEE_Signal_Process_Mag_2012}
{G. Hinton, L. Deng, D. Yu, G. E. Dahl, A. Mohamed, N. Jaitly, A. Senior, V.
  Vanhoucke, P. Nguyen, T. Sainath, and B. Kingsbury}, ``Deep neural networks
  for acoustic modeling in speech recognition: The shared views of four
  research groups,'' \emph{IEEE Signal Processing Magazine}, vol.~29, no.~6,
  Nov. 2012.

\bibitem{T_Sainath_IEEETran_2017_1}
{T. Sainath, R. J. Weiss, K. W. Wilson, B. Li, A. Narayanan, E. Variani, M.
  Bacchiani, I. Shafran, A. Senior, K. Chin, A. Misra, and {\chanwcom}},
  ``{Multichannel signal processing with deep neural networks for automatic
  speech recognition},'' \emph{IEEE/ACM Trans. Audio, Speech, Lang. Process.},
  Feb. 2017.

\bibitem{S_Hochreiter_neural_computation_1997_00}
S.~Hochreiter and J.~Schmidhuber, ``Long short-term memory,'' \emph{Neural
  Computation}, no.~9, pp. 1735--1780, Nov. 1997.

\bibitem{B_Li_INTERSPEECH_2017_1}
{{B. Li, T. Sainath, A. Narayanan, J. Caroselli, M. Bacchiani, A. Misra, I.
  Shafran, H. Sak, G. Pundak, K. Chin, K-C Sim, R. Weiss, K. Wilson, E.
  Variani, {\chanwcom}, O. Siohan, M. Weintraub, E. McDermott, R. Rose, and M.
  Shannon}}, ``{Acoustic modeling for Google Home},'' in
  \emph{INTERSPEECH-2017}, Aug. 2017, pp. {399--403}.

\bibitem{samsung_bixby}
``Samsung bixby,'' \url{http://www.samsung.com/bixby/}.

\bibitem{w_chan_icassp_2016_00}
W.~{Chan}, N.~{Jaitly}, Q.~{Le}, and O.~{Vinyals}, ``Listen, attend and spell:
  A neural network for large vocabulary conversational speech recognition,'' in
  \emph{2016 IEEE International Conference on Acoustics, Speech and Signal
  Processing (ICASSP)}, March 2016, pp. 4960--4964.

\bibitem{r_prabhavalkar_interspeech_2017_00}
\BIBentryALTinterwordspacing
R.~Prabhavalkar, K.~Rao, T.~N. Sainath, B.~Li, L.~Johnson, and N.~Jaitly, ``A
  comparison of sequence-to-sequence models for speech recognition,'' in
  \emph{Proc. Interspeech 2017}, 2017, pp. 939--943. [Online]. Available:
  \url{http://dx.doi.org/10.21437/Interspeech.2017-233}
\BIBentrySTDinterwordspacing

\bibitem{j_chorowski_nips_2015_00}
\BIBentryALTinterwordspacing
J.~K. Chorowski, D.~Bahdanau, D.~Serdyuk, K.~Cho, and Y.~Bengio,
  ``Attention-based models for speech recognition,'' in \emph{Advances in
  Neural Information Processing Systems 28}, C.~Cortes, N.~D. Lawrence, D.~D.
  Lee, M.~Sugiyama, and R.~Garnett, Eds.\hskip 1em plus 0.5em minus 0.4em\relax
  Curran Associates, Inc., 2015, pp. 577--585. [Online]. Available:
  \url{http://papers.nips.cc/paper/5847-attention-based-models-for-speech-recognition.pdf}
\BIBentrySTDinterwordspacing

\bibitem{c_chiu_icassp_2018_00}
C.-C. {Chiu}, T.~N. {Sainath}, Y.~{Wu}, R.~{Prabhavalkar}, P.~{Nguyen},
  Z.~{Chen}, A.~{Kannan}, R.~J. {Weiss}, K.~{Rao}, E.~{Gonina}, N.~{Jaitly},
  B.~{Li}, J.~{Chorowski}, and M.~{Bacchiani}, ``State-of-the-art speech
  recognition with sequence-to-sequence models,'' in \emph{2018 IEEE
  International Conference on Acoustics, Speech and Signal Processing
  (ICASSP)}, April 2018, pp. 4774--4778.

\bibitem{r_Prabhavalkar_icassp_2018_00}
R.~{Prabhavalkar}, T.~N. {Sainath}, Y.~{Wu}, P.~{Nguyen}, Z.~{Chen}, C.~{Chiu},
  and A.~{Kannan}, ``Minimum word error rate training for attention-based
  sequence-to-sequence models,'' in \emph{2018 IEEE International Conference on
  Acoustics, Speech and Signal Processing (ICASSP)}, April 2018, pp.
  4839--4843.

\bibitem{a_narayanan_slt_2018_00}
A.~{Narayanan}, A.~{Misra}, K.~C. {Sim}, G.~{Pundak}, A.~{Tripathi},
  M.~{Elfeky}, P.~{Haghani}, T.~{Strohman}, and M.~{Bacchiani}, ``Toward
  domain-invariant speech recognition via large scale training,'' in \emph{2018
  IEEE Spoken Language Technology Workshop (SLT)}, Dec 2018, pp. 441--447.

\bibitem{h_soltau_interspeech_2017_00}
\BIBentryALTinterwordspacing
H.~Soltau, H.~Liao, and H.~Sak, ``Neural speech recognizer: Acoustic-to-word
  lstm model for large vocabulary speech recognition,'' in
  \emph{INTERSPEECH-2017}, 2017, pp. 3707--3711. [Online]. Available:
  \url{http://dx.doi.org/10.21437/Interspeech.2017-1566}
\BIBentrySTDinterwordspacing

\bibitem{E_Variani_INTERSPEECH_2017_01}
\BIBentryALTinterwordspacing
{E. Variani, T. Bagby, E. McDermott, and M. Bacchiani}, ``{End-to-end training
  of acoustic models for large vocabulary continuous speech recognition with
  tensorflow},'' in \emph{{INTERSPEECH-2017}}, 2017, pp. 1641--1645. [Online].
  Available: \url{http://dx.doi.org/10.21437/Interspeech.2017-1284}
\BIBentrySTDinterwordspacing

\bibitem{p_goyal_arxiv_2018_00}
\BIBentryALTinterwordspacing
P.~Goyal, P.~Doll{\'{a}}r, R.~B. Girshick, P.~Noordhuis, L.~Wesolowski,
  A.~Kyrola, A.~Tulloch, Y.~Jia, and K.~He, ``Accurate, large minibatch {SGD:}
  training imagenet in 1 hour,'' \emph{CoRR}, vol. abs/1706.02677, 2017.
  [Online]. Available: \url{http://arxiv.org/abs/1706.02677}
\BIBentrySTDinterwordspacing

\bibitem{C_Kim_IEEETran_2016_1}
{\chanwcom} and R.~M. Stern, ``{Power-Normalized Cepstral Coefficients (PNCC)
  for Robust Speech Recognition},'' \emph{IEEE/ACM Trans. Audio, Speech, Lang.
  Process.}, pp. 1315--1329, July 2016.

\bibitem{U_H_Yapanel_SpeechComm_2008}
{U. H. Yapanel and J. H. L. Hansen}, ``{A new perceptually motivated MVDR-based
  acoustic front-end (PMVDR) for robust automatic speech recognition},''
  \emph{{Speech Communication}}, vol.~50, no.~2, pp. 142--152, {Feb.} 2008.

\bibitem{c_kim_interspeech_2014_00}
{\chanwcom}, K.~{Chin}, M.~{Bacchiani}, and R.~M. {Stern}, ``Robust speech
  recognition using temporal masking and thresholding algorithma,'' in
  \emph{INTERSPEECH-2014}, Sept. 2014, pp. 2734--2738.

\bibitem{T_Nekatani_ICASSP_2017_1}
{T. Nakatani, N. Ito, T. Higuchi, S. Araki, and K. Kinoshita}, ``{Integrating
  DNN-based and spatial clustering-based mask estimation for robust MVDR
  beamforming},'' in \emph{IEEE Int. Conf. Acoust., Speech, Signal Processing},
  March 2017, pp. 286--290.

\bibitem{T_Higuchi_ICASSP_2016_1}
{T. Higuchi and N. Ito and T. Yoshioka and T. Nakatani}, ``{Robust MVDR
  beamforming using time-frequency masks for online/offline ASR in noise},'' in
  \emph{IEEE Int. Conf. Acoust., Speech, Signal Processing}, March 2016, pp.
  5210--5214.

\bibitem{H_Erdogan_INTERSPEECH_2016_1}
{H. Erdogan, J. R. Hershey, S. Watanabe, M. Mandel, J. Roux}, ``{Improved MVDR
  Beamforming Using Single-Channel Mask Prediction Networks},'' in
  \emph{INTERSPEECH-2016}, Sept 2016, pp. 1981--1985.

\bibitem{C_Kim_INTERSPEECH_2010_1}
{{\chanwcom}, K. Eom, J. Lee, and R. M. Stern}, ``Automatic selection of
  thresholds for signal separation algorithms based on interaural delay,'' in
  \emph{INTERSPEECH-2010}, Sept. 2010, pp. 729--732.

\bibitem{C_Kim_ICASSP_2012_2}
{{\chanwcom}, C. Khawand, and R. M. Stern}, ``Two-microphone source separation
  algorithm based on statistical modeling of angle distributions,'' in
  \emph{IEEE Int. Conf. on Acoustics, Speech, and Signal Processing}, March
  2012, pp. 4629--4632.

\bibitem{C_Kim_INTERSPEECH_2015_1}
{{\chanwcom} and K. K. Chin}, ``Sound source separation algorithm using phase
  difference and angle distribution modeling near the target,'' in
  \emph{INTERSPEECH-2015}, Sept. 2015, pp. 751--755.

\bibitem{c_kim_icassp_2018_01}
{\chanwcom}, A.~{Menon}, M.~{Bacchiani}, and R.~{Stern}, ``Sound source
  separation using phase difference and reliable mask selection selection,'' in
  \emph{2018 IEEE International Conference on Acoustics, Speech and Signal
  Processing (ICASSP)}, April 2018, pp. 5559--5563.

\bibitem{R_Lippmann_icassp_1987_1}
{R. Lippmann, E. Martin, and D. Paul}, ``Multi-style training for robust
  isolated-word speech recognition,'' in \emph{IEEE International Conference on
  Acoustics, Speech, and Signal Processing}, vol.~12, Apr 1987, pp. 705--708.

\bibitem{c_kim_icassp_2018_00}
{\chanwcom}, T.~{Sainath}, A.~{Narayanan}, A.~{Misra}, R.~{Nongpiur}, and
  M.~{Bacchiani}, ``Spectral distortion model for training phase-sensitive
  deep-neural networks for far-field speech recognition,'' in \emph{2018 IEEE
  International Conference on Acoustics, Speech and Signal Processing
  (ICASSP)}, April 2018, pp. 5729--5733.

\bibitem{w_hartmann_interspeech_2016_00}
\BIBentryALTinterwordspacing
W.~Hartmann, T.~Ng, R.~Hsiao, S.~Tsakalidis, and R.~Schwartz, ``Two-stage data
  augmentation for low-resourced speech recognition,'' in
  \emph{INTERSPEECH-2016}, 2016, pp. 2378--2382. [Online]. Available:
  \url{http://dx.doi.org/10.21437/Interspeech.2016-1386}
\BIBentrySTDinterwordspacing

\bibitem{s_park_interspeech_2019_00}
\BIBentryALTinterwordspacing
D.~S. Park, W.~Chan, Y.~Zhang, C.-C. Chiu, B.~Zoph, E.~D. Cubuk, and Q.~V. Le,
  ``{SpecAugment: A Simple Data Augmentation Method for Automatic Speech
  Recognition},'' in \emph{Proc. Interspeech 2019}, 2019, pp. 2613--2617.
  [Online]. Available: \url{http://dx.doi.org/10.21437/Interspeech.2019-2680}
\BIBentrySTDinterwordspacing

\bibitem{C_Kim_ASRU_2009_2}
{{\chanwcom}, K. Kumar and R. M. Stern}, ``Robust speech recognition using
  small power boosting algorithm,'' in \emph{IEEE Automatic Speech Recognition
  and Understanding Workshop}, {Dec.} 2009, pp. 243--248.

\bibitem{c_kim_interspeech_2018_00}
\BIBentryALTinterwordspacing
{\chanwcom}, E.~Variani, A.~Narayanan, and M.~Bacchiani, ``Efficient
  implementation of the room simulator for training deep neural network
  acoustic models,'' in \emph{INTERSPEECH-2018}, Sept 2018, pp. 3028--3032.
  [Online]. Available: \url{http://dx.doi.org/10.21437/Interspeech.2018-2566}
\BIBentrySTDinterwordspacing

\bibitem{n_jaitly_icml_workshop_2013_00}
N.~Jaitly and G.~E. Hinton, ``Vocal tract length perturbation (vtlp) improves
  speech recognition,'' in \emph{Int. Conf. Mach. Learn. (ICML) Workshop on
  Deep Learn. Audio, Speech, Lang. Process.}, 2013.

\bibitem{m_abadi_usenix_2016}
\BIBentryALTinterwordspacing
M.~Abadi, P.~Barham, J.~Chen, Z.~Chen, A.~Davis, J.~Dean, M.~Devin,
  S.~Ghemawat, G.~Irving, M.~Isard, M.~Kudlur, J.~Levenberg, R.~Monga,
  S.~Moore, D.~G. Murray, B.~Steiner, P.~Tucker, V.~Vasudevan, P.~Warden,
  M.~Wicke, Y.~Yu, and X.~Zheng, ``Tensorflow: A system for large-scale machine
  learning,'' in \emph{12th {USENIX} Symposium on Operating Systems Design and
  Implementation ({OSDI} 16)}.\hskip 1em plus 0.5em minus 0.4em\relax Savannah,
  GA: {USENIX} Association, 2016, pp. 265--283. [Online]. Available:
  \url{https://www.usenix.org/conference/osdi16/technical-sessions/presentation/abadi}
\BIBentrySTDinterwordspacing

\bibitem{a_sergeev_arxiv_2018_00}
A.~Sergeev and M.~D. Balso, ``Horovod: fast and easy distributed deep learning
  in {TensorFlow},'' \emph{arXiv preprint arXiv:1802.05799}, 2018.

\bibitem{ibm_spectrum_lsf_2010}
IBM, \emph{{IBM Spectrum LSF, Version 10 Release 1.0, Configuration
  Reference}}.

\bibitem{x_cui_taslp_2015_00}
X.~Cui, V.~Goel, and B.~Kingsbury, ``Data augmentation for deep neural network
  acoustic modeling,'' \emph{IEEE/ACM Transactions on Audio, Speech, and
  Language Processing}, vol.~23, no.~9, pp. 1469--1477, Sept 2015.

\bibitem{p_zhan_cmu_tech_report_1997_00}
\BIBentryALTinterwordspacing
{P. Zhan and A. Waibel}, ``Vocal tract length normalization for large
  vocabulary continuous speech recognition,'' School of Computer Science,
  Carnegie Mellon University, Tech. Rep. CMU-CS-97-148, May 1997. [Online].
  Available:
  \url{https://www.lti.cs.cmu.edu/sites/default/files/CMU-LTI-97-150-T.pdf}
\BIBentrySTDinterwordspacing

\bibitem{c_kim_asru_2019_00}
{\chanwcom}, M.~Kumar, K.~Kim, and D.~Gowda, ``Power-law nonlinearity with
  maximally uniform distribution criterion for improved neural network training
  in automatic speech recognition,'' in \emph{2019 IEEE Automatic Speech
  Recognition and Understanding Workshop (ASRU)}, Dec. 2019 (accepted).

\bibitem{C_Kim_ICASSP_2010_1}
{{\chanwcom} and R. M. Stern}, ``Feature extraction for robust speech
  recognition based on maximizing the sharpness of the power distribution and
  on power flooring,'' in \emph{IEEE Int. Conf. on Acoustics, Speech, and
  Signal Processing}, March 2010, pp. 4574--4577.

\bibitem{C_Kim_ICASSP_2012_1}
------, ``Power-normalized cepstral coefficients (pncc) for robust speech
  recognition,'' in \emph{IEEE Int. Conf. on Acoustics, Speech, and Signal
  Processing}, March 2012, pp. 4101--4104.

\bibitem{C_Kim_PhDThesis_2010}
{\chanwcom}, ``Signal processing for robust speech recognition motivated by
  auditory processing,'' Ph.D. dissertation, Carnegie Mellon University,
  Pittsburgh, PA USA, Dec. 2010.

\bibitem{C_Kim_INTERSPEECH_2009_2}
{{\chanwcom} and R. M. Stern}, ``Feature extraction for robust speech
  recognition using a power-law nonlinearity and power-bias subtraction,'' in
  \emph{INTERSPEECH-2009}, {Sept.} 2009, pp. 28--31.

\bibitem{zero_mq}
``Zero mq,'' \url{http:zeromq.org}.

\bibitem{c_chiu_iclr_2018_00}
\BIBentryALTinterwordspacing
C.-C. Chiu and C.~Raffel, ``Monotonic chunkwise attention,'' in
  \emph{International Conference on Learning Representations}, Apr. 2018.
  [Online]. Available: \url{https://openreview.net/forum?id=Hko85plCW}
\BIBentrySTDinterwordspacing

\bibitem{k_kim_asru_2019_00}
K.~Kim, K.~Lee, D.~Gowda, J.~Park, S.~Kim, S.~Jin, Y.-Y. Lee, J.~Yeo, D.~Kim,
  S.~Jung, J.~Lee, M.~Han, and {\chanwcom}, ``attention based on-device
  streaming speech recognition with large speech corpus,'' in \emph{2019 IEEE
  Automatic Speech Recognition and Understanding Workshop (ASRU)}, Dec. 2019
  (accepted).

\bibitem{p_doetsch_icassp_2017_00}
P.~Doetsch, A.~Zeyer, P.~Voigtlaender, I.~Kulikov, R.~Schl{\"u}ter, and H.~Ney,
  ``{RETURNN: the RWTH extensible training framework for universal recurrent
  neural networks},'' in \emph{IEEE Int. Conf. Acoust., Speech, and Signal
  Processing}, March 2017, pp. 5345--5349.

\bibitem{a_zeyer_interspeech_2018_00}
\BIBentryALTinterwordspacing
A.~Zeyer, K.~Irie, R.~Schl{\"u}ter, and H.~Ney, ``{Improved training of
  end-to-end attention models for speech recognition},'' in
  \emph{INTERSPEECH-2018}, 2018, pp. 7--11. [Online]. Available:
  \url{http://dx.doi.org/10.21437/Interspeech.2018-1616}
\BIBentrySTDinterwordspacing

\bibitem{d_gowda_interspeech_2019_00}
\BIBentryALTinterwordspacing
D.~Gowda, A.~Garg, K.~Kim, M.~Kumar, and {\chanwcom}, ``Multi-task
  multi-resolution char-to-bpe cross-attention decoder for end-to-end speech
  recognition,'' in \emph{INTERSPEECH-2019}, Graz, Austria, Sept. 2019, pp.
  2783--2787. [Online]. Available:
  \url{http://dx.doi.org/10.21437/Interspeech.2019-3216}
\BIBentrySTDinterwordspacing

\bibitem{a_garg_asru_2019_00}
A.~Garg, D.~Gowda, A.~Kumar, K.~Kim, M.~Kumar, and {\chanwcom}, ``Improved
  multi-stage training of online attention-based encoder-decoder models,'' in
  \emph{2019 IEEE Automatic Speech Recognition and Understanding Workshop
  (ASRU)}, Dec. 2019 (accepted).

\bibitem{r_pascanu_icml_2013}
\BIBentryALTinterwordspacing
R.~Pascanu, T.~Mikolov, and Y.~Bengio, ``On the difficulty of training
  recurrent neural networks,'' in \emph{Proceedings of the 30th International
  Conference on International Conference on Machine Learning - Volume 28}, ser.
  ICML'13.\hskip 1em plus 0.5em minus 0.4em\relax JMLR.org, 2013, pp.
  III--1310--III--1318. [Online]. Available:
  \url{http://dl.acm.org/citation.cfm?id=3042817.3043083}
\BIBentrySTDinterwordspacing

\bibitem{c_gulcehre_corr_2015_00}
\BIBentryALTinterwordspacing
{\c{C}}.~G{\"{u}}l{\c{c}}ehre, O.~Firat, K.~Xu, K.~Cho, L.~Barrault, H.~Lin,
  F.~Bougares, H.~Schwenk, and Y.~Bengio, ``On using monolingual corpora in
  neural machine translation,'' \emph{CoRR}, vol. abs/1503.03535, 2015.
  [Online]. Available: \url{http://arxiv.org/abs/1503.03535}
\BIBentrySTDinterwordspacing

\bibitem{s_toshniwal_slt_2018_00}
S.~{Toshniwal}, A.~{Kannan}, C.~{Chiu}, Y.~{Wu}, T.~N. {Sainath}, and
  K.~{Livescu}, ``A comparison of techniques for language model integration in
  encoder-decoder speech recognition,'' in \emph{2018 IEEE Spoken Language
  Technology Workshop (SLT)}, Dec 2018, pp. 369--375.

\bibitem{b_mcfee_proc_scipy_2015_00}
B.~McFee, C.~Raffel, D.~Liang, D.~P. Ellis, M.~McVicar, E.~Battenberg, and
  O.~Nieto, ``librosa: Audio and music signal analysis in python,'' in
  \emph{{P}roceedings of the 14th {P}ython in {S}cience {C}onference}, K.~Huff
  and J.~Bergstra, Eds., 2015, pp. 18 -- 25.

\bibitem{j_heymann_icassp_2016_00}
J.~{Heymann}, L.~{Drude}, and R.~{Haeb-Umbach}, ``Neural network based spectral
  mask estimation for acoustic beamforming,'' in \emph{2016 IEEE International
  Conference on Acoustics, Speech and Signal Processing (ICASSP)}, March 2016,
  pp. 196--200.

\bibitem{a_vaswani_nips_2017_00}
\BIBentryALTinterwordspacing
A.~Vaswani, N.~Shazeer, N.~Parmar, J.~Uszkoreit, L.~Jones, A.~N. Gomez, L.~u.
  Kaiser, and I.~Polosukhin, ``Attention is all you need,'' in \emph{Advances
  in Neural Information Processing Systems 30}, I.~Guyon, U.~V. Luxburg,
  S.~Bengio, H.~Wallach, R.~Fergus, S.~Vishwanathan, and R.~Garnett, Eds.\hskip
  1em plus 0.5em minus 0.4em\relax Curran Associates, Inc., 2017, pp.
  5998--6008. [Online]. Available:
  \url{http://papers.nips.cc/paper/7181-attention-is-all-you-need.pdf}
\BIBentrySTDinterwordspacing

\bibitem{c_luscher_arxiv_2019_00}
\BIBentryALTinterwordspacing
C.~L{\"{u}}scher, E.~Beck, K.~Irie, M.~Kitza, W.~Michel, A.~Zeyer,
  R.~Schl{\"{u}}ter, and H.~Ney, ``{RWTH} {ASR} systems for librispeech: Hybrid
  vs attention - w/o data augmentation,'' \emph{CoRR}, vol. abs/1905.03072,
  2019. [Online]. Available: \url{http://arxiv.org/abs/1905.03072}
\BIBentrySTDinterwordspacing

\end{thebibliography}

\end{document}